\documentclass[12pt]{iopart}
\usepackage[dvips]{graphicx}
\usepackage{cite}
\usepackage{iopams}
\usepackage{subfigure}
\usepackage{wasysym}
\usepackage{bbold}

\newcommand{\rd}{{\rm d}}
\newcommand{\re}{{\rm e}}
\newcommand{\ri}{{\rm i}}

\newcommand{\be}{\begin{equation}}
\newcommand{\ee}{\end{equation}}
\newcommand{\bea}{\begin{eqnarray}}
\newcommand{\eea}{\end{eqnarray}}

\newcommand{\nn}{\nonumber}

\newcommand{\ket}[1]{|#1\rangle} 
 
\newcommand{\eka}{\left[}
\newcommand{\ekz}{\right]}
\newcommand{\rka}{\left(}
\newcommand{\rkz}{\right)}

\begin{document}

\title{Manipulation of matter waves using Bloch and Bloch-Zener oscillations}

\author{B. M. Breid, D. Witthaut and H. J. Korsch}

\address{FB Physik, Technische Universit\"at Kaiserslautern,
D--67653 Kaiserslautern, Germany}

\ead{korsch@physik.uni-kl.de}

\begin{abstract}

\noindent
We present theoretical and numerical results on the dynamics of ultracold atoms
in an accelerated single- and double-periodic optical lattice. In the
single-periodic potential Bloch oscillations can be used to generate fast 
directed transport with very little dispersion. The dynamics in the
double-periodic system is dominated by Bloch-Zener oscillations, i.e. the interplay 
of Bloch oscillations and Zener tunneling between the subbands.
Apart from directed transport, the latter system permits various interesting
applications, such as widely tunable matter wave beam splitters and Mach-Zehnder
interferometry. As an application, a method for efficient probing of small nonlinear
mean-field interactions is suggested.
\end{abstract}


\maketitle

\section{Introduction}
\label{intro}
The experimental progress in storing and controlling ultracold
atoms in optical lattices (see, e.g. \cite{Eier03, Mors06}) has led to a variety of spectacular
results in the last decade, as for instance the superfluid to Mott insulator
phase transition \cite{Grei02}.
Also the field of linear and nonlinear atom optics has
benefited a lot from cooling and storing atoms in optical lattices.
Early results include the observation of Bloch oscillations in
accelerated lattices \cite{Daha96, Peik97} and coherent pulsed output from
a BEC in a vertical lattice under the influence of gravity \cite{Ande98}.
Today it is a matter of routine to prepare a wave packet in a state
of well defined quasi momentum by accelerating the optical lattice.

Combining Bloch oscillations and Zener tunneling between Bloch
bands offers new possibilities to control the dynamics
of cold atoms.
However, in a usual cosine-shaped optical potential, the band gaps
decrease rapidly with increasing energy. A matter wave packet
tunneling from the ground band to the first excited band will therefore
also tunnel to even higher bands and finally escape to infinity. Indeed
this happens, e.g., in the Kasevich experiment \cite{Ande98,02pulse}.
However, systems can be constructed that avoid decay and still allow
Zener tunneling between certain (mini)bands. In fact, this can be
achieved by introducing a second, double-periodic optical potential.
This leads to a splitting of the ground band into two minibands that
are still energetically well separated from all excited bands. A matter
wave packet under the influence of an external field will Bloch
oscillate, whereas Zener tunneling between the minibands will lead to a
splitting of the wave packet and to interference.  

In this paper we investigate the dynamics of cold atoms in a one-dimensional
double-peridic potential, which is governed by the Schr\"odinger equation
\be
  \rmi\hbar\frac{\partial}{\partial t} \Psi(x,t) =
  \eka-\frac{\hbar^2}{2m}\frac{\partial^2}{\partial x^2}
  + U \cos \frac{2\pi x}{d} + \varepsilon U \cos \frac{\pi x}{d} + F(t)x\ekz \Psi(x,t) \,
  \label{G5_90}
\ee
Here, $d$ denotes the fundamental period, $F$ is the strength of the external
field and $U$ and $\varepsilon U$ are the amplitudes of the two optical lattices,
where the double-periodic potential is weak, $\varepsilon \ll 1$.
For convenience we use scaled units such that the fundamental
period is $d_s = 2\pi$ and the amplitude of the deeper lattice is $U_s = 1$,
\be
  x_s = \frac{2\pi}{d}\,x \, , \quad
  \hbar_s = \frac{2\pi}{d\sqrt{Um}}\,\hbar \, , \quad
  F_s(t) = \frac{d}{2\pi U}\,F(t) \, , \quad
  t_s = \frac{2\pi}{d}\sqrt{\frac{U}{m}}\, t \, ,
\ee
which leads to the dimensionless Schr\"odinger equation
\be
  \rmi\hbar_s\frac{\partial}{\partial t_s} \Psi(x,t) =
  \eka-\frac{\hbar_s^2}{2}\frac{\partial^2}{\partial x_s^2}
  +\cos\rka x_s\rkz + \varepsilon \cos\rka\frac{x_s}{2}\rkz +
  F_s(t) x_s\ekz \Psi(x,t) \, . \label{G5_96}
\ee
Unless otherwise stated, the parameter values are chosen as
$\hbar_s = 2.828$ and $F_s = 0.0011$, which corresponds to the
experimental setup of the Arimondo group \cite{Cris04} for $\varepsilon=0$.
These are typical experimental dimensions. 
In contrast to the cited experiments, we are not going to take inter-atomic
interactions into account except for section \ref{probing}.
This is achieved experimentally by a very low density of particles.
In the following we will omit the index $s$ to simplify notation.

This paper is organized as follows:
We start by reviewing some important results for the single-periodic
potential, i.e. $\varepsilon = 0$, in section \ref{sec-single-per}.
Furthermore we discuss in section \ref{sec-shuttle} a shuttling
mechanism for transporting wave packets in optical lattices by flipping
the external field.
It is shown that the transport velocity is independent of the field strength and that dispersion is negligible.
The case of a double-periodic potential is then discussed in section
\ref{sec-double-per}, starting with a brief description of the dynamics
of Bloch-Zener oscillations.
Combining this effect with the shuttling transport mechanism
offers the possibility to construct a highly controllable
matter wave beam splitter as described in section \ref{K5_3_3}.
Finally we discuss the possibility of matter wave Mach-Zehnder
interferometry and a possible application for probing small
mean-field interactions in Bose-Einstein condensates in section
\ref{mzi} and \ref{probing}.

\section{Single-periodic potentials}
\label{sec-single-per}

\subsection{Bloch oscillations}

Bloch oscillations of quantum particles in periodic potentials under
the influence of a static external field $F(t)=F$ have been predicted
already in 1928 \cite{Bloc28}.
The recent experimental progress with cold atoms in optical lattices has
triggered a renewed theoretical interest in this topic (see
\cite{04bloch1d,04bloch2d,04bloch} for recent reviews).

\begin{figure}[htb]
\centering
\includegraphics[height=6cm, angle=0]{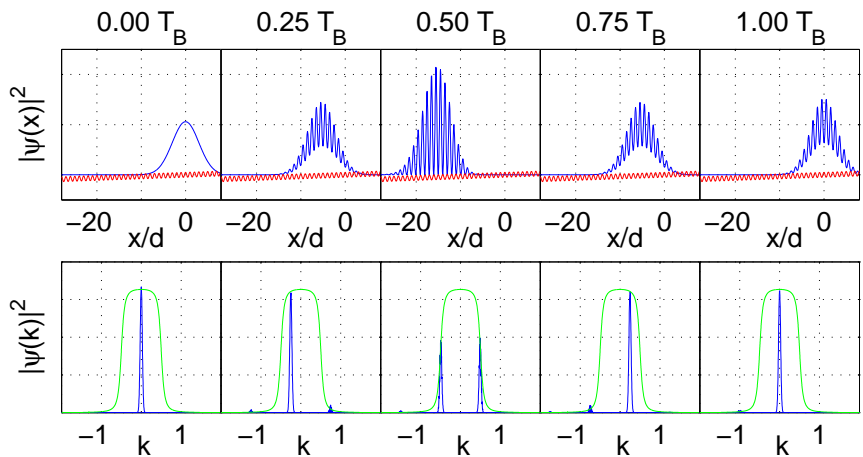}
\caption{\label{5_21}
Schematic evolution of a Bloch oscillation of a gaussian wave packet in
position space (top) and in momentum space (bottom) with $\varepsilon=0$
(compare \cite{03bloch2D}).}
\end{figure}

The dynamics of Bloch oscillations is illustrated in figure \ref{5_21}
in real space and in momentum space.
A handwaving explanation of these oscillations can be given easily, 
assuming that the external field tilts the energy of the Bloch bands 
in real space. Quite often, all higher bands are energetically far from the
ground band and the Bloch oscillation can already be understood within
a single-band approximation.
A wave packet in the ground band is accelerated in real space by the 
external field and reflected at the edges of the band, which gives rise to an
oscillating motion. Within this picture, the spatial extension of these oscillations
can be estimated as
\be
   L = \frac{\Delta}{F} \, ,
   \label{eqn-bloch-disp}
\ee
where $\Delta$ denotes the energy width of the Bloch band. The
oscillation
period is given by the characteristic Bloch time
\be
 T_B = \frac{2\pi \hbar}{dF} \, .
\ee

However, a rigorous calculation shows that the Bloch bands do not exist
any
longer for $F \ne 0$. The spectrum of the rescaled Wannier-Stark Hamiltonian
\be
 \hat H_{WS}=-\frac{\hbar^2}{2}\frac{\partial^2}{\partial x^2} + V(x)+ Fx  
   \,,\quad V(x+d)=V(x)
\ee
is continuous with embedded resonances, the so-called Wannier-Stark
ladder of
resonances \cite{Avro77,02wsrep}. The corresponding eigenvalues are
arranged
in ladders $E_{\alpha,n}=E_{\alpha,0}+ n d F$, where $\alpha$ denotes
the ladder
index and $n$ denotes the site index. The eigenstates within one ladder
are
related by a spatial translation
$\psi_{\alpha,n}(x)=\psi_{\alpha,0}(x- nd)$,
resp. $\psi_{\alpha,n}(k)=\rme^{-\rmi  n d k  }\,\psi_{\alpha,0}(k)$
in
momentum space.

The dynamics of Bloch oscillations is now readily understood in the
Wannier-Stark
eigenbasis \cite{04bloch1d}. For weak fields, the dynamics takes place
almost
exclusively in the lowest ladder $\alpha=0$. The dynamics of an initial
wave packet $\Psi(k,0)=\sum\nolimits_n c_{\alpha=0,n}\,
\psi_{\alpha=0,n}(k)$ is then given by
\bea
\Psi(k,t)&=&\sum_n c_{0,n} \,\rme^{-\rmi E_{0,n} t/\hbar}
\,\psi_{0,n}(k) \nn \\ 
 &=&\sum_{n} c_{0,n} \,\rme^{-\rmi (E_{0,0}+ ndF)t/\hbar}\,
\rme^{-\rmi n d k } \,\psi_{0,0}(k) \nn \\ 
 &=&\rme^{-\rmi E_{0,0} t/\hbar}\, \psi_{0,0}(k) C(k+Ft/\hbar)
\eea
The function $C(k+Ft/\hbar)$ is the discrete Fourier transformation of
the
coefficients $c_{0,n}$ evaluated at the point $k+Ft/\hbar$. For a broad
initial wave packet it is a $2 \pi/d$-periodic series of narrow peaks.
The dynamics in momentum space shown in figure \ref{5_21} is now easily
understood: The function $C(k)$ moves under an envelope given by the
Wannier-Stark function $\psi_{0,0}(k)$. In real space, this periodic
motion yields the familiar Bloch oscillations.

\begin{figure}[h]
\centering
\includegraphics[height=6cm, angle=0]{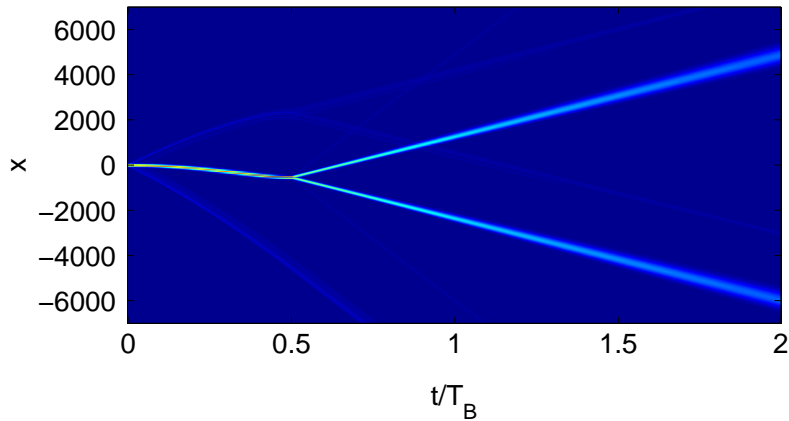}
\caption{\label{5_22}
Splitting of a gaussian wave packet in position space as described in
the text. Shown is $|\Psi(x,t)|$ as a colormap plot for
$|\varepsilon|=0$.}
\end{figure}

In view of the introduction of efficient matter-wave beam
splitters in
double-periodic potentials in section \ref{K5_3_3}, we also discuss a
very simple
mechanism of beam splitting using Bloch oscillations.
At the time $t=T_B/2$, when a wave packet with average initial quasi
momentum
$\kappa=0$ is just crossing the edge of the Brillouin zone, this wave
packet
consists of two fractions with opposite momentum (see figure
\ref{5_21}).
In case of completely switching off the periodic potential and also the
Stark
field at $t=T_B/2$, the two fractions move into opposite directions
according
to their momentum (see figure \ref{5_22}). The main disadvantage of this
method
is the strong dispersion of the free wave packet. Furthermore, the
splitted
wave packet is no longer located in a periodic potential, what is often
wanted for the further  experiments. Switching on the potential again
would
cause even stronger dispersion. In contrast, the splitting of a wave
packet
within a periodic potential can be done easily and with only little loss by
a
Bloch-Zener oscillation as will be shown in section \ref{K5_3_3}.

\subsection{Shuttling Transport}
\label{sec-shuttle}

\begin{figure}[htb]
\centering
\includegraphics[width=8cm,angle=0]{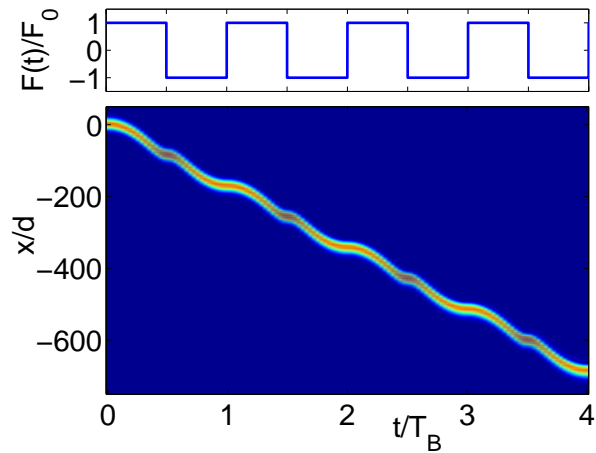}
\caption{\label{fig-bshuttle}
Shuttling transport of a gaussian wave packet in a flipped field.
Shown is $|\Psi(x,t)|$ as a colormap plot (lower panel).
Parameters are the same as in section \ref{intro}, except
that the external field $F(t)$ is periodically flipped as 
illustrated in the upper panel.}
\end{figure}

It is well known that a time-dependend external field $F(t)$ may
lead to transport or dynamical localization \cite{04bloch1d, 03TBalg}.
An effective way of transporting a wave packet with low loss in an optical lattice is
the `Bloch shuttle'
\be
  F(t) = \left\{\begin{array}{c c l}
  + F_0 & \mbox{for} & \mbox{mod}(t,T_B) \in [0,T_B/2) \\
  - F_0 & \mbox{for} & \mbox{mod}(t,T_B) \in [T_B/2,T_B) \,.
  \end{array} \right. 
\ee
The basic idea is simple. Within half of a Bloch period $T_B/2$,
the wave packet will be displaced by $L = \Delta/F_0$ and will
return to its initial position within the next half period as shown in figure \ref{5_21}.
However, if the direction of the external field is flipped, also
the direction of motion flips.      
This shuttling transport is illustrated in figure \ref{fig-bshuttle},
where the modulus of the wave function $|\Psi(x,t)|$ is plotted
for such an alternating external field (cf. \cite{04bloch1d}).
The simulation has been done with Hamiltonian (\ref{G5_96}) which takes all Bloch bands in to account. 
Note that anyway the dynamics considered here is based on Bloch oscillations which take place within the ground band.
In that sense, the transport mechanism presented here is a single-band effect. 
Since all other bands are energetically separated, their influence on the dynamics can be neglected. 
A stronger influence of higher bands due to a stronger external force would increase the dispersion.

This transport mechanism, denoted as `shuttling transport' or just
`Bloch shuttle' in the following, has some remarkable features that
can be analyzed in the single-band tight-binding approximation.
First of all, the transport velocity $v_{\rm trans}$ is {\it independent} of the field
strength $F_0$, which can be concluded easily.
Within half of the Bloch period, the wave packet is displaced by
$L = \Delta/F_0$ (cf. equation \ref{eqn-bloch-disp}), which leads to the estimate
\be
  v_{\rm trans} = \frac{L}{T_B/2} = \frac{\Delta/F_0}{\pi\hbar/dF_0}
  = \frac{d \Delta}{\pi \hbar} \, .
  \label{eqn-transp-v}
\ee
Secondly, as observed in figure \ref{fig-bshuttle}, the width of the wave packet is
nearly conserved. No dispersion can be detected in this figure.
In fact one can prove within a tight-binding model, that the width  
\be
\Delta^2_x(t)=\langle x^2\rangle_t-\langle x\rangle^2_t
\ee
of an initially broad gaussian wave packet
is in leading order given by 
\be
\Delta^2_x(n T_B)-\Delta_x^2(0)\approx \frac{\Delta^2d^4}{8F_0^2} \, \frac{n^2}{\sigma_x^{4}} \, 
\ee
(see appendix). The dispersion vanishes rapidly with increasing spatial width $\sigma_x$ of the initial wave packet. 
For comparison, a free gaussian wave packet of a particle with mass $m$ spreads as
\be
\Delta^2_x(t)-\Delta_x^2(0)= \frac{\hbar^2}{4m^2} \, \frac{t^2}{\sigma_x^{2}} \, .
\ee
In conclusion dispersion is negligible for all relevant applications.

Note that this transport mechanism resembles the shuttling transport of
electric charge in electro-mechanical Coulomb blockade nanostructures
\cite{Gore98}.
Here, a metallic grain can vibrate between two leads with a voltage
drop $V$. Near the turning points of the oscillation electrons tunnel
onto the grain (in the vicinity of the negative lead) or off the grain
(in the vicinity of the positive lead).
Thus, the grain is alternately negatively or positively charged and thus
accelerated in the electrical field between the leads.
Several aspects of the Coulomb shuttle are in close analogy to the Bloch
shuttle discussed above.
The system switches periodically between two states: negatively/positively
charged corresponding to the field directions $F(t) = \pm F_0$.
Depending on this state, the system performs a half oscillation into one
direction or the other. Due to the periodical recharging, respectively the flipping
of the field, the resulting transport is always directed into the same
direction. In both realizations the net transport does {\it not} depend
on the voltage $V$ respectively the external field $F_0$.
However, considering the nanoscale electron shuttle, the current may depend
implicitly on $V$ through the oscillation frequency or the number of tunneling
electrons.
The analogy is not total, of course. The voltage drop leads to a preferred
transport direction through the nano device, whereas the transport direction
of a Bloch shuttle for cold atoms is determined only by the initial field
direction $F(t = 0)$.

\section{Double-periodic potentials}
\label{sec-double-per}

\subsection{Bloch-Zener oscillations}
\label{B-Z-O}

In this section we study the coherent superposition of Bloch oscillations
as described in section \ref{sec-single-per} and Zener tunneling, denoted as Bloch-Zener oscillations.
The dynamics is given by the scaled Schr\"odinger equation (\ref{G5_96})
with a double-periodic potential, i.e. $\varepsilon \ne 0$.

\begin{figure}[htb]
\centering
\includegraphics[width=7.7cm,height=5 cm,  angle=0]{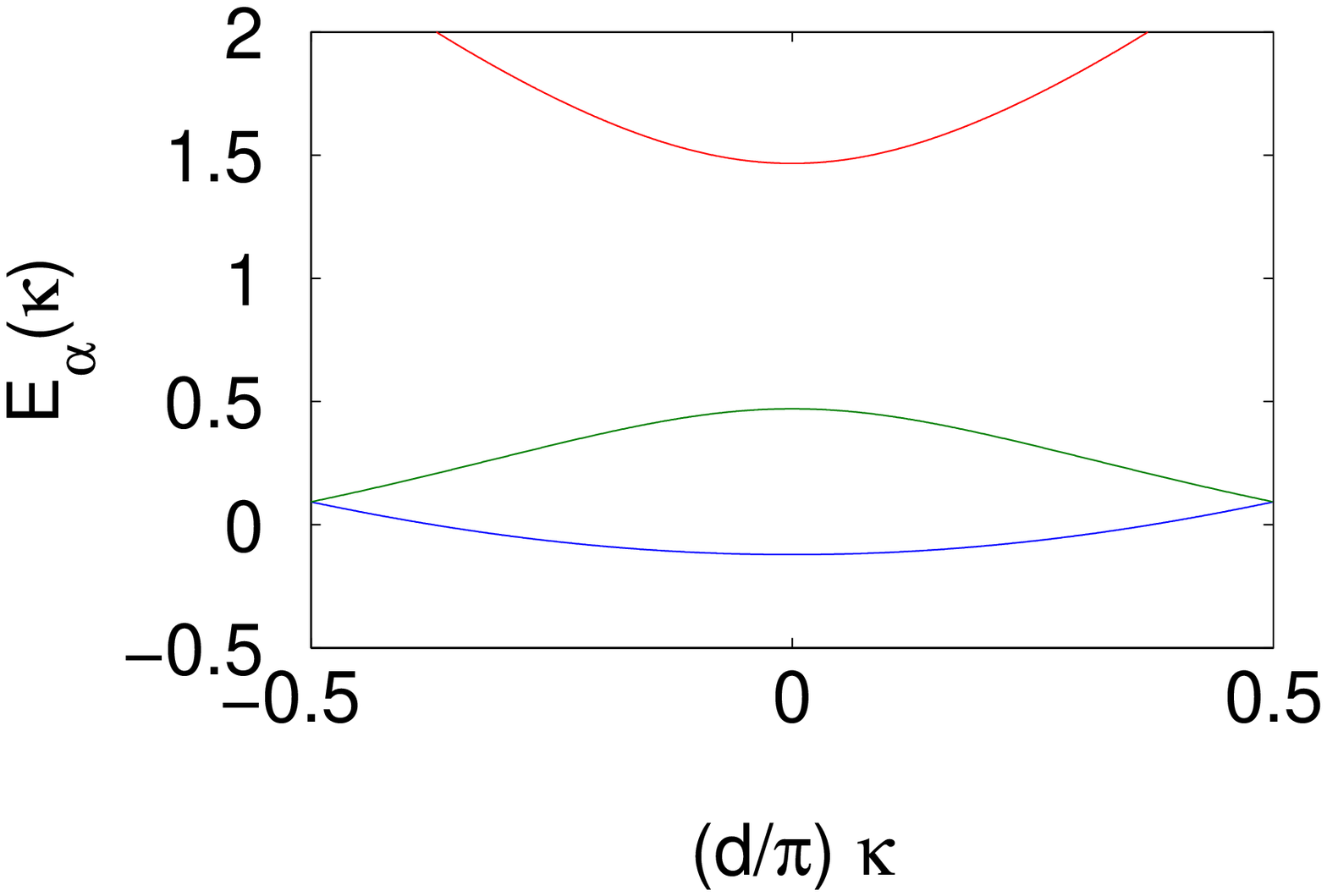}
\includegraphics[width=7.7cm,height=5 cm,  angle=0]{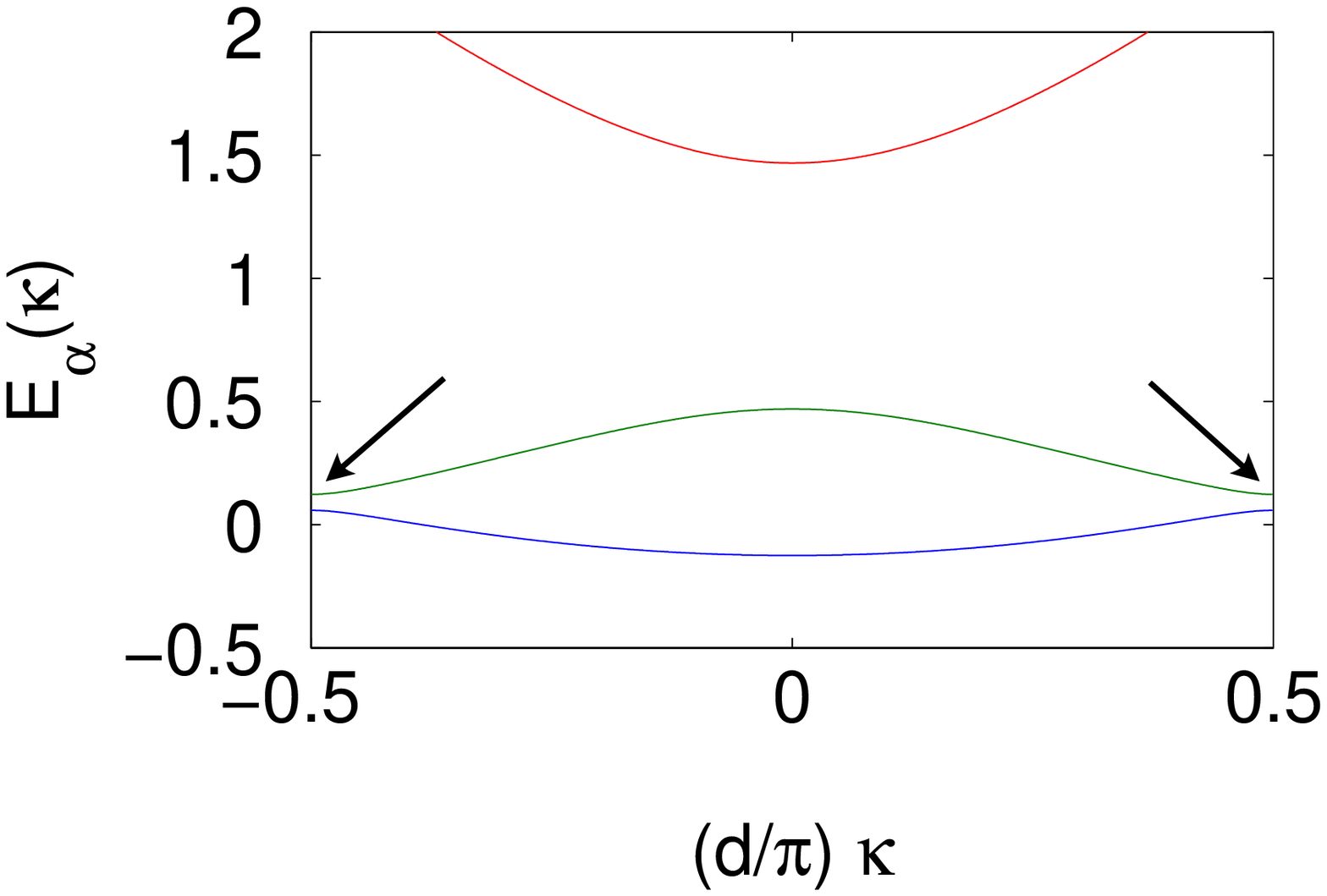}
\caption{\label{Zusatz}
Dispersion relation of the system (\ref{G5_96}) with the parameters $\hbar=2.828$, $F=0$. Left: $\varepsilon=0$ (bands `folded' into the reduced Brillouin zone). Right: $\varepsilon=0.121$.}
\end{figure}

In the field-free case, the additional double-periodic potential leads to a
splitting of the ground Bloch band into two minibands. This is shown in figure
\ref{Zusatz}, where the dispersion relation for $\varepsilon=0.121$ is compared
to the one of the single-periodic system $\varepsilon=0$. Because of the large energy gap between the two minibands and the next higher band, the dynamics of the system is mainly affected by these minibands. Therefore, the dynamics is expected to be similar to the two-band tight-binding model (see \cite{06bloch_zener}). 

In the presence of a constant external field $F$, the spectrum of (\ref{G5_96}) consists of Wannier-Stark ladders
instead of Bloch bands \cite{02wsrep}. Due to the double-periodic potential, even these ladders
split up into two `miniladders' just like the Bloch bands of the field free system.
This splitting was rigorously proven for a corresponding two-band tight-binding system
recently \cite{06bloch_zener}, where it was also shown that the parameters
of the tight-binding system can be chosen in a way to reach periodic reconstruction
of an arbitrary initial wave packet. We expect to find a similar behaviour for the more realistic potential considered here.

In order to discuss the general featurs of the dynamics of Bloch-Zener oscillations,
we expand a given wave packet in the Wannier-Stark eigenstates of the system (\ref{G5_90}),
\be
  \hat H \ket{\psi_{\alpha,n}} = E_{\alpha,n} \ket{\psi_{\alpha,n}} \,,
\ee
where $\alpha=0,1$ denotes the miniladder index and $n$ denotes the site index. 
For weak external fields $F$, decay can be neglected and only the two miniladders corresponding to the lowest minibands (the two Wannier-Stark resonances with the least decay) have to be taken into account. The two energy ladders can be written as
\be
  E_{0,n} = E_0 + 2ndF \quad \mbox{and} \quad E_{1,n} = E_1 + 2ndF
\ee
where $E_0$ and $E_1$ are the energy offsets of the two Wannier-Stark miniladders. 
The eigenstates with different site indices are related by a
spatial translation 
\be
|\psi_{\alpha,n}(x)\rangle = |\psi_{\alpha,0}(x-2nd)\rangle \,.
\ee
Now, an arbitrary initial wave packet $\ket{\Psi(t=0)}$ can be expanded
in the Wannier-Stark basis,
\be
  \ket{\Psi(t=0)} = \sum_n c_{\alpha=0,n} \ket{\psi_{\alpha=0,n}(x)}
   + \sum_n c_{\alpha=1,n} \ket{\psi_{\alpha=1,n}(x)}.
\ee
Writing the Wannier-Stark states in the Bloch basis and using the phase change
\be
\ket{\chi_{0,\kappa}(x+2d)}=\rme^{\rmi 2d\kappa}\ket{\chi_{0,\kappa}(x)} \quad\mbox{and}\quad
\ket{\chi_{1,\kappa}(x+2d)}=\rme^{\rmi 2d\kappa}\ket{\chi_{1,\kappa}(x)} \, 
\ee
of the Bloch waves under spatial translations, the time evolution of the wave packet $\ket{\Psi(t)}$ is given by 
\bea
\fl 
\hspace{0.8cm}
  \quad &&\langle \chi_{0,\kappa}\ket{\Psi(t)} =
  \rme^{-\frac{\rmi}{\hbar}E_0t}\eka a_{0,0}(\kappa)\,C_0\rka\textstyle\kappa+\frac{Ft}{\hbar}\displaystyle\rkz
  +a_{1,0}(\kappa)\,\rme^{-\frac{\rmi}{\hbar}(E_1-E_0)t}C_1\rka\textstyle\kappa+\frac{Ft}{\hbar}\displaystyle\rkz\ekz
  \label{G5_31} \, ,\\
\fl 
\hspace{0.8cm}  
  \quad && \langle \chi_{1,\kappa}\ket{\Psi(t)} =
  \rme^{-\frac{\rmi}{\hbar}E_0t}\eka b_{0,0}(\kappa)\,C_0\rka\textstyle\kappa+\frac{Ft}{\hbar}\displaystyle\rkz
  +b_{1,0}(\kappa)\,\rme^{-\frac{\rmi}{\hbar}(E_1-E_0)t}C_1\rka\textstyle\kappa+\frac{Ft}{\hbar}\displaystyle\rkz\ekz \, ,
  \label{G5_32}
\eea
where $\ket{\chi_{0,\kappa}}$ and $\ket{\chi_{1,\kappa}}$ are the Bloch waves ($F=0$) of the ground and the first excited miniband. 
The functions $C_0\rka\kappa\rkz$ and  $C_1\rka\kappa\rkz$
are the discrete Fourier transforms of the expansion coefficients $c_{0,n}$ and
$c_{1,n}$, respectively, and the functions $a_{0,0}(\kappa)$, $a_{1,0}(\kappa)$, $b_{0,0}(\kappa)$
and $b_{1,0}(\kappa)$ are the coefficients of the Wannier-Stark functions in the $\kappa$-basis.
Note that all six functions are $\pi/d$-periodic in $\kappa$. These equations are similar to the two-band tight-binding model \cite{06bloch_zener}.

This result, which is an extension of the corresponding tight-binding equations \cite{06bloch_zener}, leads to some interesting effects.
The dynamics of the two band system is characterized by two time scales. The functions
$C_0(\kappa)$ and $C_1(\kappa)$ are reconstructed at multiples of
\be
T_1=\frac{\pi \hbar}{d F} \,,\label{t1}
\ee
whereas the exponential function $\rme^{-\frac{\rmi}{\hbar}(E_1-E_0)t}$ has a period of
\be
T_2=\frac{2\pi \hbar}{E_1-E_0} \, . \label{t2}
\ee
The period $T_1$ is just half of the Bloch time $T_B=\frac{2\pi \hbar}{d F}$ of the single-periodic system, $\varepsilon=0$, which we take as a reference time in the following. If a wave function consists only of states of a single energy ladder, one of the functions $C_0$, $C_1$ is zero for all times $t$. In this case the initial state is reconstructed up to a global phase after a period $T_1$ which is just an ordinary Bloch oscillation.

Whenever the commensurability condition
\be
\frac{T_1}{T_2}=\frac{E_1-E_0}{2dF}=\frac{s}{r} \quad\mbox{with}\quad r,s\in\mathbb{N}
\ee
is fulfilled, the functions (\ref{G5_31}) and (\ref{G5_32}) reconstruct at multiples of the
Bloch-Zener time
\be
  T_{BZ}=sT_2=rT_1
\ee
up to a global phase shift.
Furthermore the dynamics of the occupation probability of the two minibands at multiples
of the time $T_1$ can be expressed in the form (cf. \cite{06bloch_zener})
\bea
&&\int\limits_{-\frac{\pi}{2d}}^{\frac{\pi}{2d}}|\langle \chi_{0,\kappa}\ket{\Psi(nT_1)}|^2\,\rmd\kappa=X+Y\cos\rka\frac{E_1-E_0}{dF}\,\pi\, n+\varphi\rkz \label{G5_71} \, ,\\
&&\int\limits_{-\frac{\pi}{2d}}^{\frac{\pi}{2d}}|\langle \chi_{1,\kappa}\ket{\Psi(nT_1)}|^2\,\rmd\kappa=1-\eka X+Y\cos\rka\frac{E_1-E_0}{dF}\,\pi\, n+\varphi\rkz\ekz \,, \label{G5_72}
\eea
where $X$ and $Y$ are real positive numbers. This follows from the equations (\ref{G5_31}),\,(\ref{G5_32}) by a straightforward
calculation. 
In the case of $T_2$ and $T_1$ being commensurate, these equations read
\bea
&&\int\limits_{-\frac{\pi}{2d}}^{\frac{\pi}{2d}}|\langle \chi_{0,\kappa}\ket{\Psi(nT_1)}|^2\,\rmd\kappa=X+Y\cos\rka2\pi\,\frac{s}{r}\, n+\varphi\rkz 
\label{G5_74} \, ,\\
&&\int\limits_{-\frac{\pi}{2d}}^{\frac{\pi}{2d}}|\langle \chi_{1,\kappa}\ket{\Psi(nT_1)}|^2\,\rmd\kappa=1-\eka X+Y\cos\rka2\pi\,\frac{s}{r}\, n+\varphi\rkz\ekz \, ,
\label{G5_75}
\eea 
where one recognizes the complete reconstruction at multiples of $T_{BZ}$.
In the case of $T_2$ and $T_1$ being incommensurate, $\varphi=0$ holds
whenever only one of the bands is initially occupied.

\begin{figure}[htb]
\centering
\includegraphics[width=7.7cm, height=5cm,  angle=0]{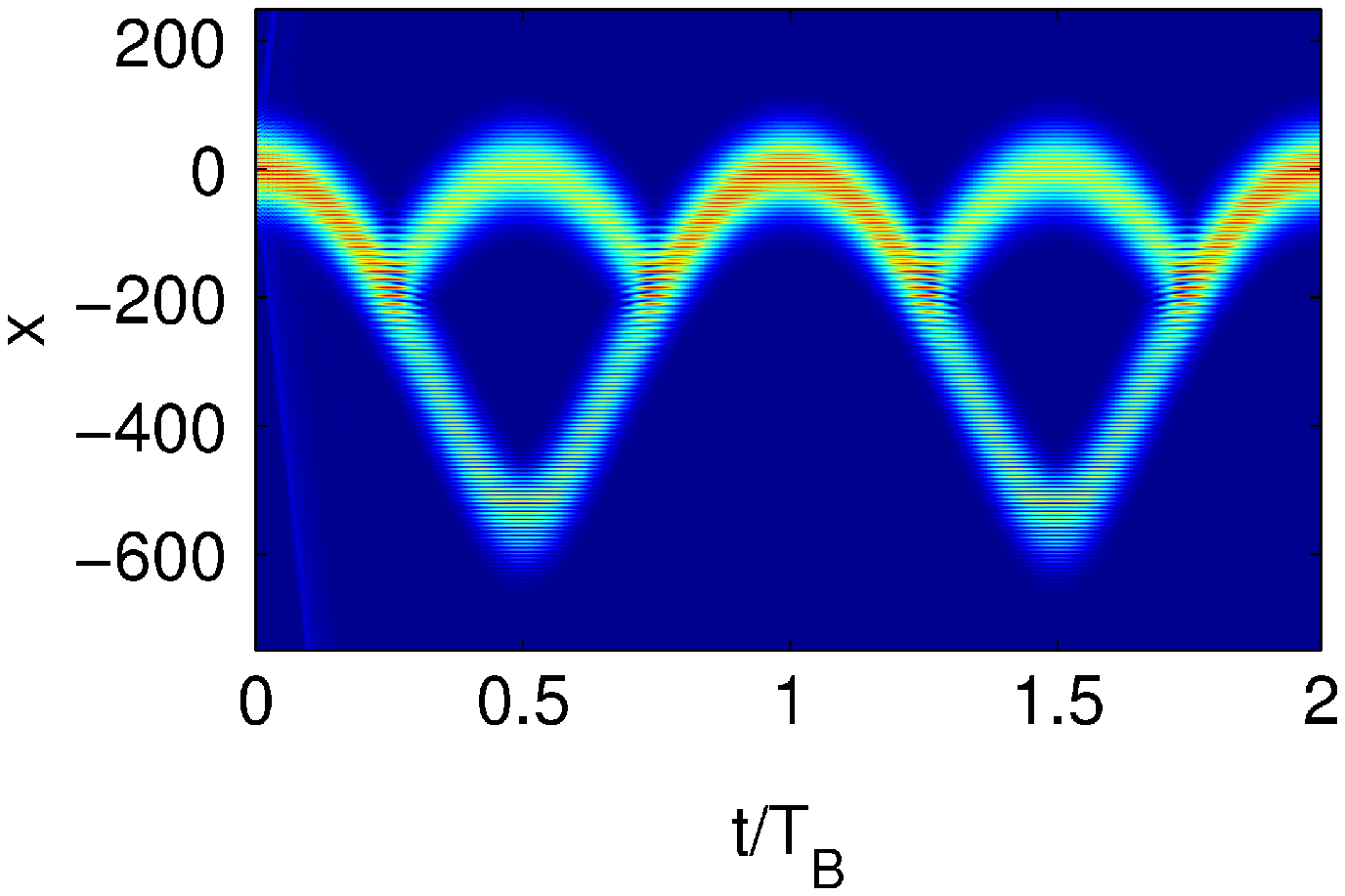}
\includegraphics[width=7.7cm, height=5cm,  angle=0]{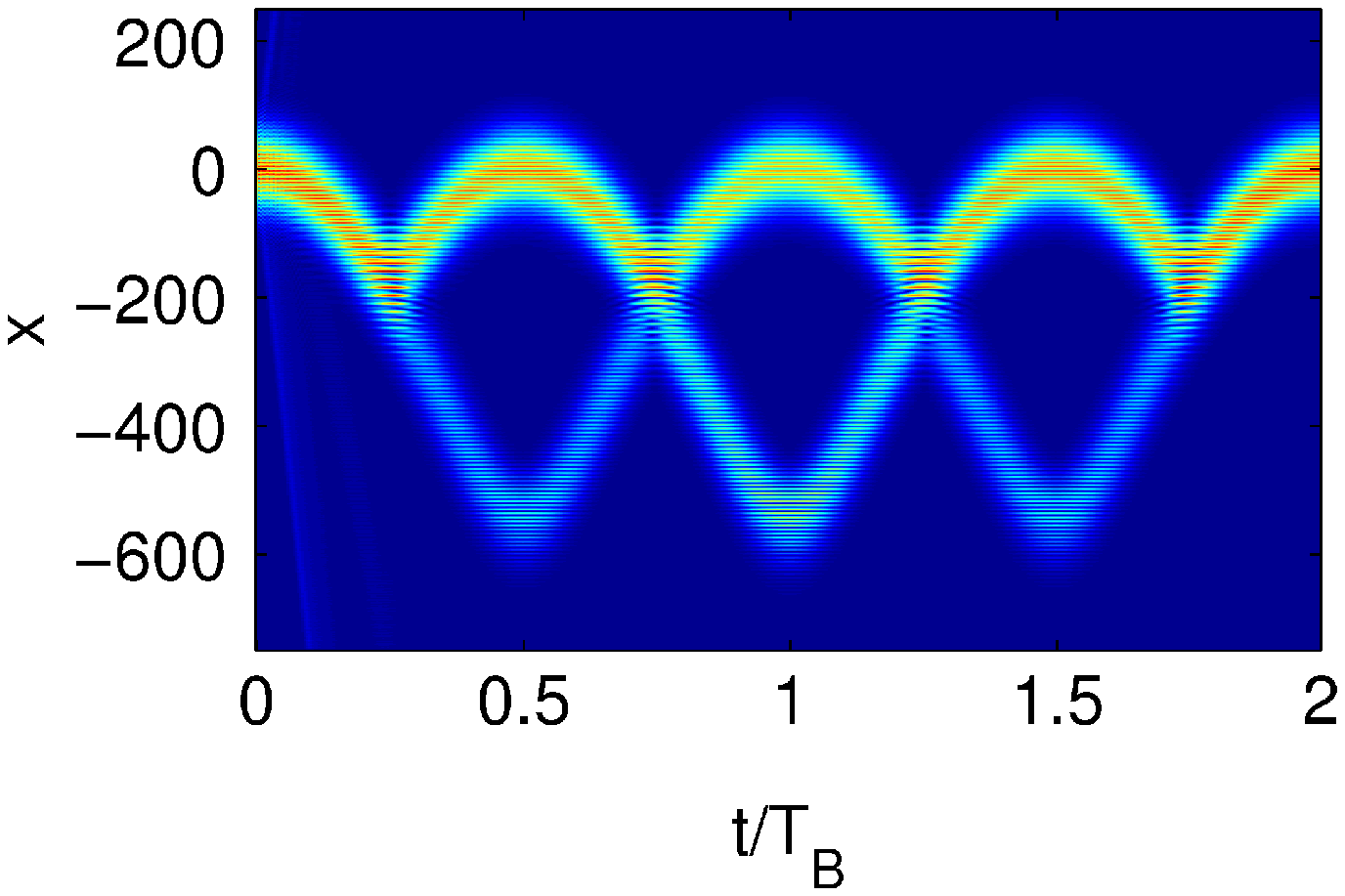}
\caption{\label{5_14} \label{5_15}
Time evolution of an initially gaussian wave packet.
Shown is $|\Psi(x,t)|$ as a colormap plot for $\varepsilon=0.0825$ (left)
and for $\varepsilon=0.121$ (right).}
\end{figure}

We now present some numerical results for the dynamics given by the `realistic' 
Schr\"odinger equation (\ref{G5_96}) with a period-doubled potential.
The time evolution has been calculated using a split-operator method
\cite{Feit82}, where the initial state is a real ($\kappa_0=0$) normalized gaussian
\be
  \Psi(t=0,x) \sim\rme^{-\rka\frac{x}{60}\rkz^2}
\ee
throughout this section.
Figure \ref{5_14} shows the results for two different values of
$\varepsilon$. The time $t$ is given in units of the Bloch time $T_B$
of the single-band system $\varepsilon=0$.

In order to understand the dynamics of the wave packet, it is instructive to consider the time evolution in quasimomentum space.
Remember that the motion of a wave packet under the influence of a constant force $F$ in quasimomentum space (cf. figure \ref{Zusatz}) follows the acceleration theorem
\be
\kappa(t)=\kappa_0-{Ft}/{\hbar} \,. \label{at}
\ee
The main part of the wave packet shows a superposition of Bloch
oscillations and Zener tunneling between the two minibands, the
Bloch-Zener oscillations.
Zener tunneling takes place almost exclusively when the wavepacket
reaches the edge of the reduced Brillouin zone, where the tunneling
rate strongly depends on $\varepsilon$.
In figure \ref{5_14} one clearly sees the splitting of the wave packet
due to Zener tunneling around $t = T_B/4$. In general, the fractions
from the two bands interfere, which gives rise to splitting and
reconstruction.
The parameter $\varepsilon$ in figure \ref{5_14} is chosen in a way that
reconstruction appears after one or two Bloch
times, respectively. 

The group velocity of a wave packet is proportional 
to the slope of the dispersion relation. At the edge of the Brillouin zone 
the slope of the wave packet does not change for the fraction of the wave packet 
which tunnels into the other miniband but does so for the fraction remaining in its miniband. 
The results are two interfering oscillations of different amplitudes as shown in figure \ref{5_14}.    

The fractions of the wave packet in higher bands escape to $-\infty$ quite quickly.
Depending on the slope of the particular band (see figure \ref{Zusatz}) this happens
initially in the direction of the positive or negative $x$-axis. 

\subsection{Transport}
In section \ref{sec-shuttle} we presented a transport mechanism based on a periodic field flip. The whole process can be understood within a \textit{single-band} approximation. Even for a \textit{two-band} system one can achieve transport of a gaussian wave packet by switching the signs of some system parameters. 
In order to reach transport not only the field strength $F$ but also the amplitude of the double-periodic part of the potential in equation (\ref{G5_96}), the parameter $\varepsilon$ has to be flipped periodically. 
We can achieve transport by the sequence of parameters given in table 1.
\begin{table}[h]
\centering
\begin{tabular}{|c  | c c c | c c c | c c c | c c c| }
\hline
$t/T_B$ &   \multicolumn{2}{l}{0} 
        &   \multicolumn{2}{c}{0.5} 
	& & \multicolumn{2}{c}{1} 
	& & \multicolumn{2}{c}{1.5} 
	&   \multicolumn{2}{r|}{2}  \\ \hline
$\varepsilon $&&   $+$  & & &  $-$  & & &  $-$  & &&   $+$  & \\ \hline
$F$           &&   $+$  & & &  $+$  & & &  $-$  & &&   $-$  & \\ \hline
\end{tabular}
\caption{}
\end{table}
\begin{figure}[htb]
\centering
\includegraphics[height=6.3cm,  angle=0]{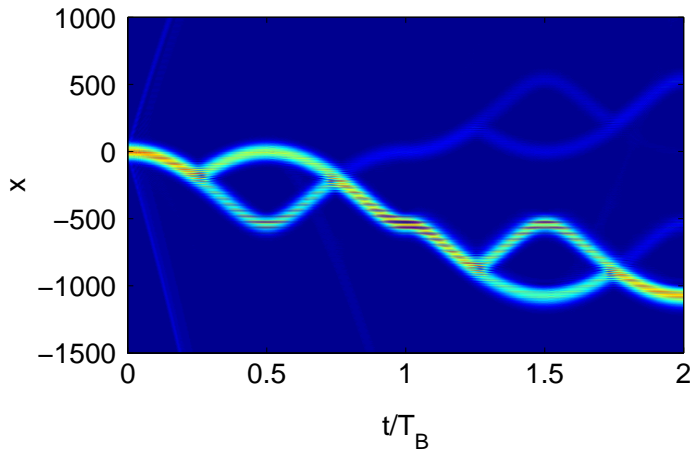}
\caption{\label{5_17} 
Transport of a gaussian wave packet. Shown is $|\Psi(x,t)|$ as a colormap plot for $|\varepsilon|=0.0825$. The parameter sequence is shown in table 1.}
\end{figure}

Apart from loss, this transport can be continued arbitrarily far by repetition of the parameter sequence.
In figure \ref{5_17} the modulus of the wave function is shown. One can easily see the fractions of the wave packet which move away from the main part at the beginning. They are caused by the occupation of higher bands as described in the previous section. Whenever the splitted wave packet interferes at the edge of the Brillouin zone, the loss increases. 

\subsection{Beam splitter}
\label{K5_3_3}
The motion of a wave packet under the influence of a constant force $F$ in quasimomentum space (cf. figure \ref{Zusatz}) follows the acceleration theorem (\ref{at}). Whenever a wave packet reaches the edge of the Brillouin zone, it can partially tunnel into the other miniband, leading to a splitting in position space (cf. figure \ref{5_14}). A permanent splitting of the wave packet can be achieved by transporting the two fractions into opposite directions. 
Thus one can realize a beam splitter in the period-doubled system (\ref{G5_96}) by applying the parameter sequence shown in table 2. The two branches of a Bloch-Zener oscillation at $t=0.5\,T_B$ are transported in opposite directions by swiching the sign of $F$ twice, once at $t=0.5\,T_B$ and once at $t=T_B$. Since the value of $\varepsilon$ is set to zero after $t=0.5\,T_B$, the transport process which separates the two branches is the same as the transport process described in section \ref{sec-shuttle}. After $t=T_B$ the field strength is constant and $\varepsilon=0$. Therefore the two wave packets continue performing ordinary Bloch oscillations.
\begin{table}[h]
\centering
\begin{tabular}{|c  | c c c | c c c | c c c | c c c| c c c|c c c|}
\hline
$t/T_B$ &   \multicolumn{2}{l}{0} 
        &   \multicolumn{2}{c}{0.5} 
	& & \multicolumn{2}{c}{1} 
	& & \multicolumn{2}{c}{1.5} 
	&   \multicolumn{2}{r|}{3}  \\ \hline
$\varepsilon $&&   $+$  & & &  $0$  & & &  $0$  & &&   $0$  & \\ \hline
$F$           &&   $+$  & & &  $-$  & & &  $+$  & &&   $+$  & \\ \hline
\end{tabular}
\caption{}
\end{table}
\begin{figure}[htb]
\centering
\includegraphics[height=6cm,  angle=0]{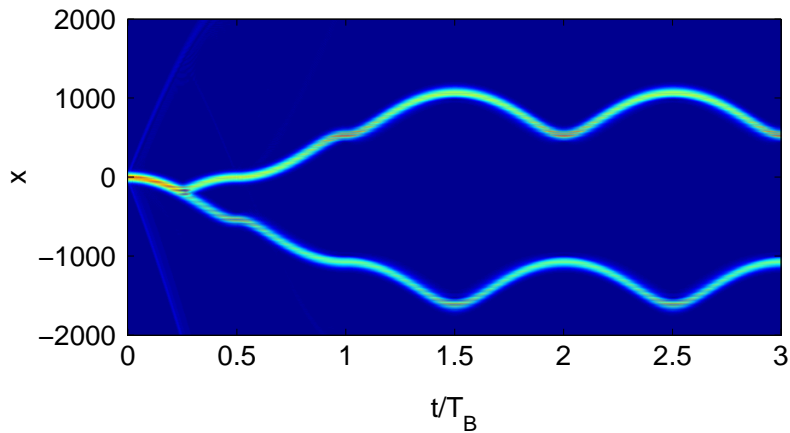}
\caption{\label{5_18} 
Splitting of a gaussian wave packet in position space. Shown is $|\Psi(x,t)|$ as a colormap plot for $|\varepsilon|=0.0825$. The parameter sequence is shown in table 2.}
\end{figure}

\begin{figure}[htb]
\centering
\includegraphics[height=6cm,  angle=0]{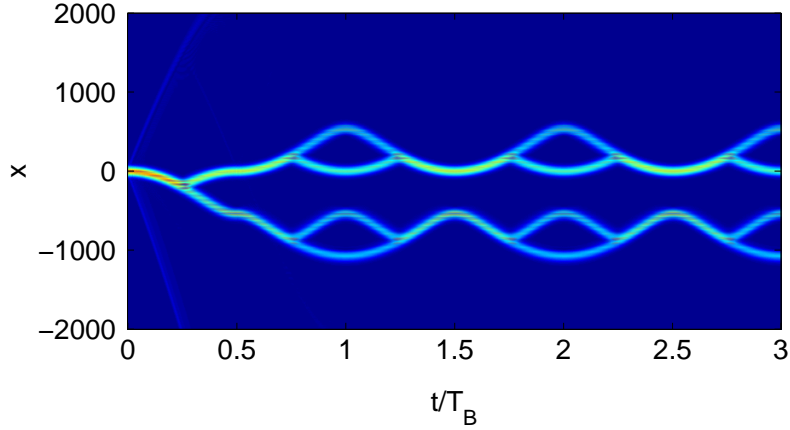}
\caption{\label{5_19} 
Splitting of a gaussian wave packet in position space. Shown is $|\Psi(x,t)|$ as a colormap plot for $|\varepsilon|=0.0825$. The parameter sequence is shown in table 3.}
\end{figure}

Another method to split the wave packet is given by the parameter sequence in table 3 which leads to the dynamics shown in figure \ref{5_19}. Here we again achieve separation of the two branches by swiching the sign of $F$. The main difference to the case above is that $\varepsilon$ is held constant during the whole process. Thus the two fractions of the splitted wave packet show Bloch-Zener oscillations instead of ordinary Bloch oscillations. 
\begin{table}[h]
\centering
\begin{tabular}{|c  | c c c | c c c | c c c | c c c| c c c|c c c|}
\hline
$t/T_B$ &   \multicolumn{2}{l}{0} 
        &   \multicolumn{2}{c}{0.5} 
	& & \multicolumn{2}{c}{1} 
	& & \multicolumn{2}{c}{1.5} 
	&   \multicolumn{2}{r|}{3}  \\ \hline
$\varepsilon $&&   $+$  & & &  $+$  & & &  $+$  & &&   $+$ & \\ \hline
$F$           &&   $+$  & & &  $-$  & & &  $-$  & &&   $-$ & \\ \hline
\end{tabular}
\caption{}
\end{table}

\begin{figure}[htb]
\centering
\includegraphics[height=6cm,  angle=0]{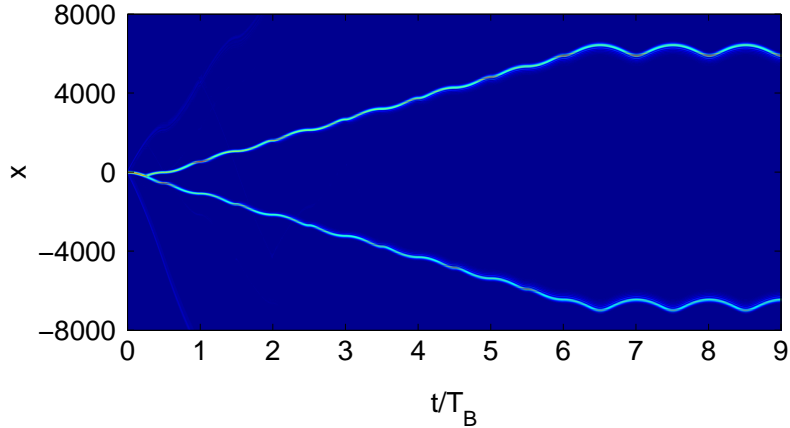}
\caption{\label{5_20} 
Splitting of a gaussian wave packet in position space. Shown is $|\Psi(x,t)|$ versus space and time with $|\varepsilon|=0.0825$. The parameter sequence is shown in table 4.}
\end{figure}

It is remarkable that the whole splitting process takes place with very little loss. Furthermore the process shown in figure \ref{5_18} can be used to separate the two fractions of the wave packet at nearly arbitrary distance. To clarify this, figure \ref{5_20} shows the dynamics of a gaussian wave packet for the parameter sequence in table 4. 
\begin{table}[h]
\centering
\begin{tabular}{|c  | c c c | c c c | c c c | c c c| c c c|c c c|}
\hline
$t/T_B$ &   \multicolumn{2}{l}{0} 
        &   \multicolumn{2}{c}{0.5} 
	& & \multicolumn{2}{c}{1} 
	& & \multicolumn{2}{c}{1.5} 
	& & \multicolumn{2}{c}{2} 
	& & \multicolumn{2}{c}{6} 
	&   \multicolumn{2}{r|}{9}  \\ \hline
$\varepsilon $&&   $+$  & & &  $0$  & & &  $0$  & &&   $0$ & &&   $0$ & &&   $0$ & \\ \hline
$F$           &&   $+$  & & &  $-$  & & &  $+$  & &&   $-$ & &&   $\pm$ alternating with $T_B/2$ & &&   $+$ & \\ \hline
\end{tabular}
\caption{}
\end{table}
Because the transport of the wave packet takes place for $\varepsilon=0$, the loss decreases strongly with an increasing width of the wave packet (see section \ref{sec-shuttle}). This is proven in the tight-binding approximation in the appendix.     

\begin{figure}[htb]
\centering
\includegraphics[height=5cm, width=7.7cm, angle=0]{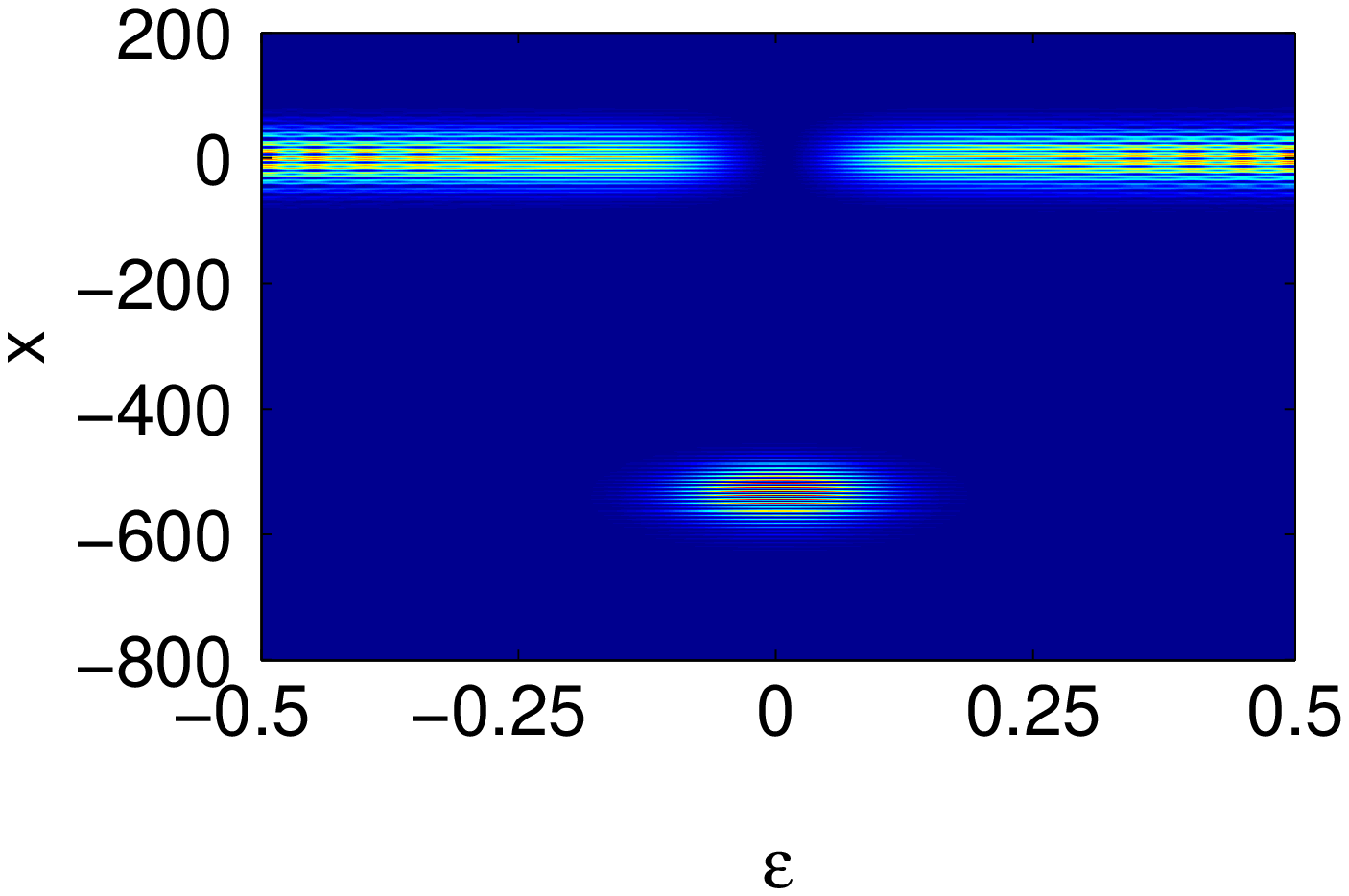}
\includegraphics[height=5cm, width=7.7cm, angle=0]{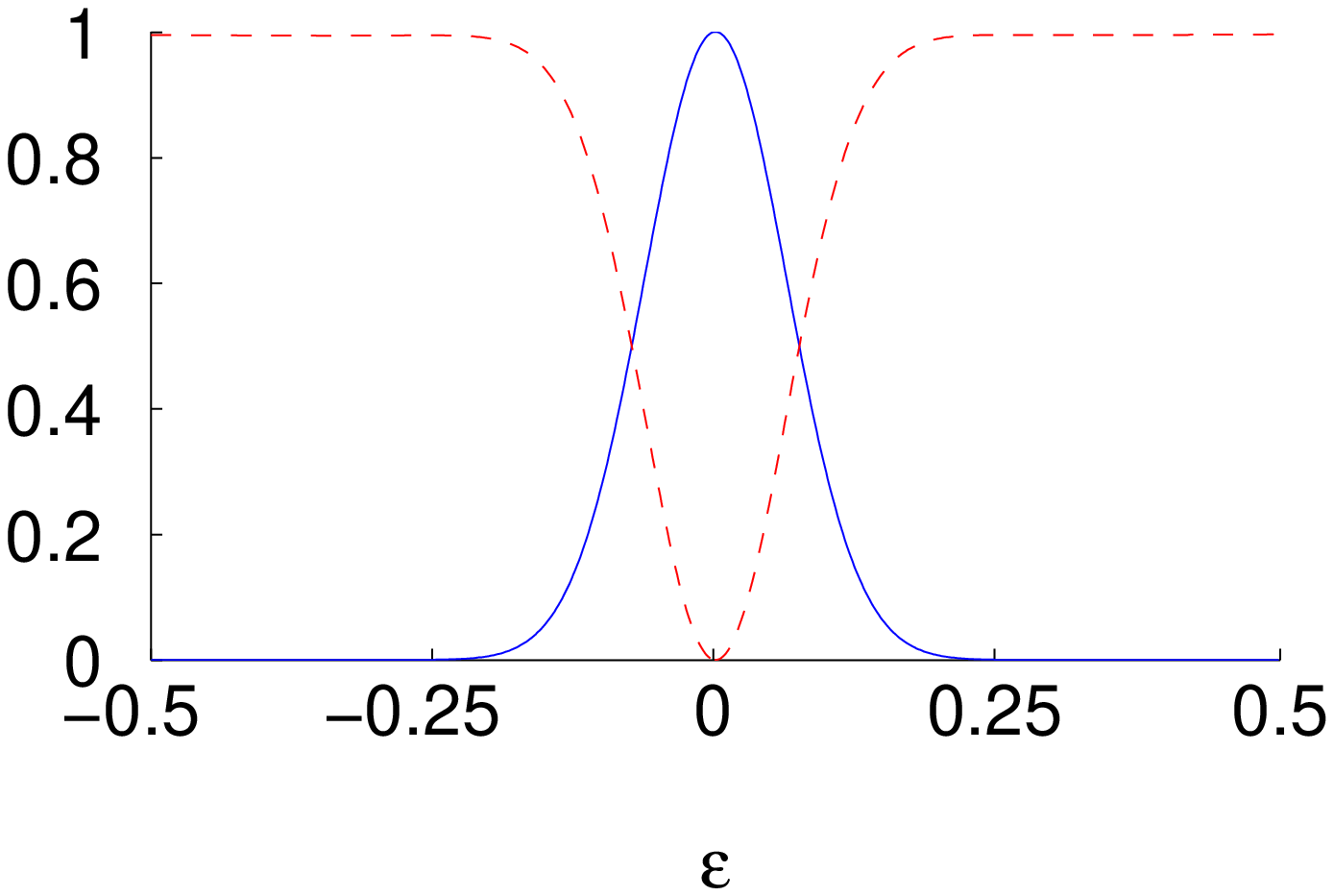}
\caption{\label{5_23} \label{5_24} 
The left-hand side shows the probability density $|\Psi(x,T_B/2)|^2$ of the wave packet after a half Bloch time $T_B/2$ versus $\varepsilon$. The right-hand side shows the integral of $|\Psi(x,T_B/2)|^2$ once over an interval $-800\leq x\leq-300$ (blue) and once over an interval $-300\leq x\leq 200$ (red, dashed), which give the occupation in the two output branches of the beam splitter, as a function of $\varepsilon$.}
\end{figure}

In addition, the control of the occupation in both branches of the splitted wave packet is quite easy by Bloch-Zener oscillations. For the splitting process analogously to figure \ref{5_18}, only the occupation in the upper and lower branch at $t=T_B/2$ is relevant. This occupation can be adjusted by the variation of $\varepsilon$ (see figure \ref{5_23}). 

In the above considerations, $\varepsilon$ was chosen in a way that the wave packet would reconstruct after a single Bloch time as long as the parameters are not switched. This choice is not mandatory. Even if $\varepsilon$ is chosen in a way that there is no reconstruction, we obtain two clearly distinguishable wave packets at $t=T_B/2$. In the picture of Bloch bands, these are just the fractions in the two different minibands. They are separated in position space by the different group velocities respectively the different slopes of the dispersion relations of the two lowest minibands. Those fractions of the wave packet, which tunnel at the edge of the Brillouin zone, will have a different group velocity than the remaining part of the wave packet. The tunneling fraction of the wave packet can be controlled by the choice of $\varepsilon$ which gives an approximate measure for the band gap. Therefore we obtain a nearly pure Bloch oscillation for small $\varepsilon$ which transports the wave packet to the lower position range. By the choice of large $\varepsilon$, the bands become separated and thus the Zener transitions are suppressed. Hence the wave packet stays in the lower miniband respectively in the upper position range.

The intermediate range $-0.2\leq \varepsilon \leq0.2$, where the occupation probability varies strongly with $\varepsilon$ is of special interest.
In this area, the occupation of both branches of the beam splitter can be adjusted very exactly (compare figure \ref{5_24}).
The distribution follows approximately the Landau-Zener formula \cite{Land32, Zene32} according to which the tunneling probability in dependence of $\varepsilon$ 
is a gaussian distribution.

In conclusion, the splitting of a wave packet within a periodic potential can be done easily and with very few loss by a Bloch-Zener oscillation as shown in figures \ref{5_18} and \ref{5_20}.
\subsection{Mach-Zehnder interferometry}
\label{mzi}
\begin{figure}[htb]
\centering
\includegraphics[height=6.5cm,  angle=0]{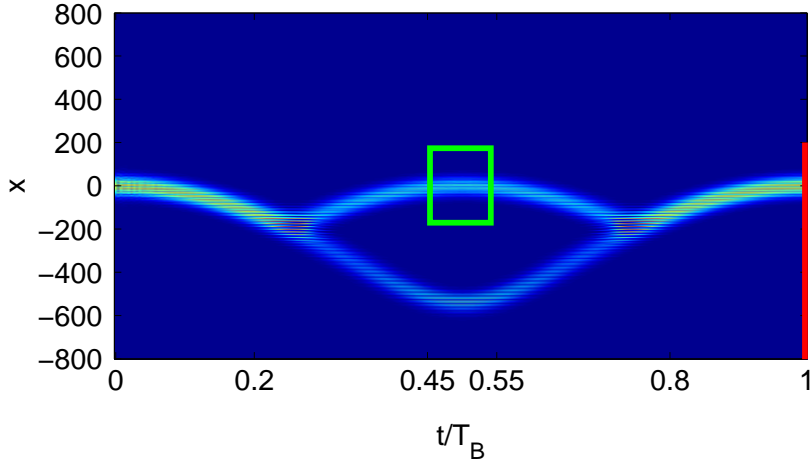}
\caption{\label{5_25} 
One branch of the splitted gaussian wave packet receives a phase shift by the additional square well potential $V_0$. Shown is $|\Psi(x,t)|^2$ as a colormap plot for $\varepsilon=0.0825$ and $V_0=0$. The position of the square well potential (green) and the probability density considered in figure \ref{5_26} at $t=T_B$ (red) are sketched.}
\end{figure}

\begin{figure}[htb]
\centering
\includegraphics[height=6.5cm,  angle=0]{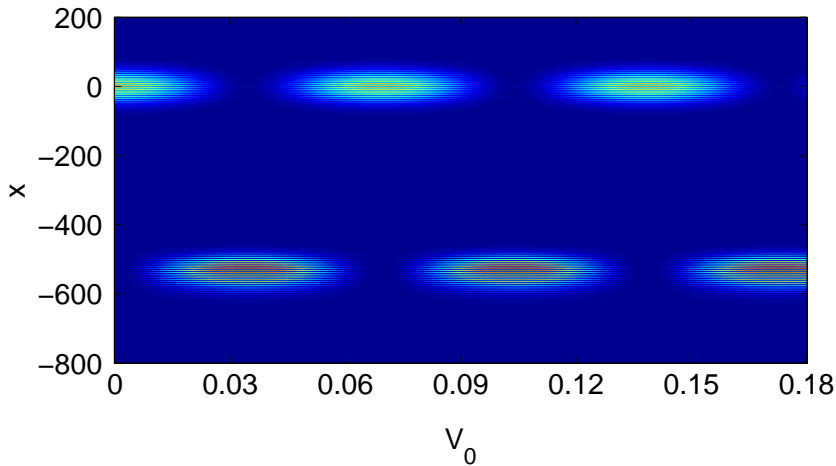}
\caption{\label{5_26} 
Shown is a colormap plot of the squared modulus of the wave function $|\Psi(x,t=T_B)|^2$ at $t=T_B$ versus the strength of the square well potential $V_0$ for $\varepsilon=0.0825$.}
\end{figure}

A very useful application of Bloch-Zener oscillations is matter wave interferometry. To this end we consider a Bloch-Zener oscillation that reconstructs after one Bloch time (see section \ref{B-Z-O}). Because the wave packet splits up into two spatially separated parts in the meantime, we can insert an additional potential into one branch as it is illustrated in figure \ref{5_25}.
Here, we apply a constant potential of strength $V_0$ in the range of $-195 \leq x\leq 195$ and within the time $0.45\,T_B\leq t \leq 0.55\,T_B$. After one Bloch time $T_B$, when both parts of the wave packet interfere again, we consider the probability density $|\Psi(x,T_B)|^2$. Figure \ref{5_26} shows the squared modulus of the wave function $|\Psi(x,T_B)|^2$ in a range of $-800 \leq x \leq 200$ at the time $t=T_B$ versus the strength of the potential $V_0$.
\begin{figure}[htb]
\centering
\includegraphics[height=6.5cm,  angle=0]{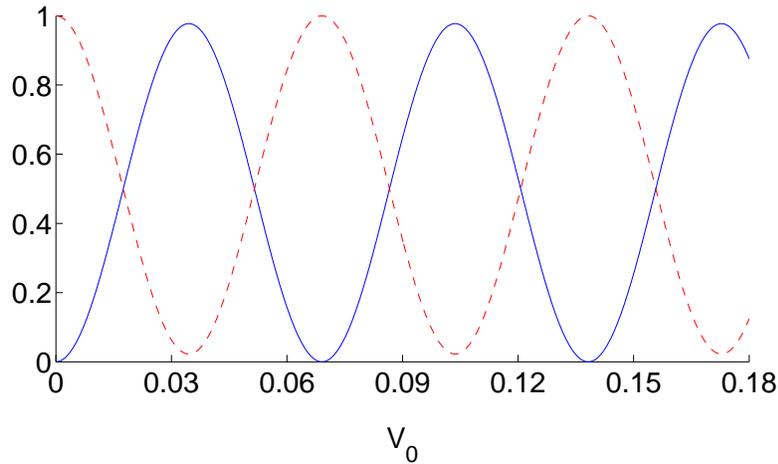}
\caption{\label{5_27} 
Integral of $|\Psi(x,t=T_B)|^2$ in figure \ref{5_26} once over a range of $-800\leq x\leq-300$ (blue) and once over a range of $-300\leq x\leq 200$ (red, dashed), giving the density in the two output branches of the interferometer.}
\end{figure}

One clearly sees that the probability distribution of the wave packet varies between the two output branches with the strength of the potential $V_0$. Depending on the phase shift that a fraction of the wave packet receives within its branch, we obtain constructive interference in the upper or the lower branch. The wave packet in the upper range is interpreted as the occupation of the lower band, the one in the lower range can be seen as the occupation of the upper band. In order to describe the interference effect more quantitatively, we integrate $|\Psi(x,T_B)|^2$ over the relevant regions. Figure \ref{5_27} shows the $V_0$-dependence of the probability to find the wave packet at $t=T_B$ in the upper and the lower region of figure \ref{5_26}. Obviously, the probability oscillates with the phase shift in the upper branch (compare figure \ref{5_25}) caused by $V_0$. The period of one oscillation is $\Delta V_0\approx 0.069$. This can be estimated by the phase shift in the upper branch
\be
\frac{\Delta V_0 \; t_{int}}{\hbar}=2\pi \quad \mbox{with} \quad t_{int}=\frac{1}{10}\,T_B=\frac{1}{10}\,\frac{\hbar}{F} \,.
\ee
Thus we get
\be
\Delta V_0=10\cdot2\pi\cdot F =10\cdot2\pi\cdot0.0011\approx 0.069 \, .
\ee
It is remarkable that the probability in the upper branch never vanishes, whereas this is the case for the probability in the lower branch. The reason is that the occupation of the interfering branches is not exactly equal. If desired, this can be achieved, however, by an adequate choice of $\varepsilon$ taken from figure \ref{5_24}. In the above examples we chose $\varepsilon$ such that the wave packet reconstructs for $V_0=0$, i.e. without an additional potential. The achieved contrast
\be
{\rm max}\eka\int_{-300}^{200}|\Psi|^2\, \rmd x \ekz_{V_0}-{\rm min}\eka\int_{-300}^{200}|\Psi|^2\, \rmd x \ekz_{V_0}\approx0.977
\ee
is very good and can even be improved by the choice of equally occupied bands. Therefore, the above method is suitable to probe weak potentials in the path of the wave packet. 

\begin{figure}[htb]
\centering
\includegraphics[height=6.5cm,  angle=0]{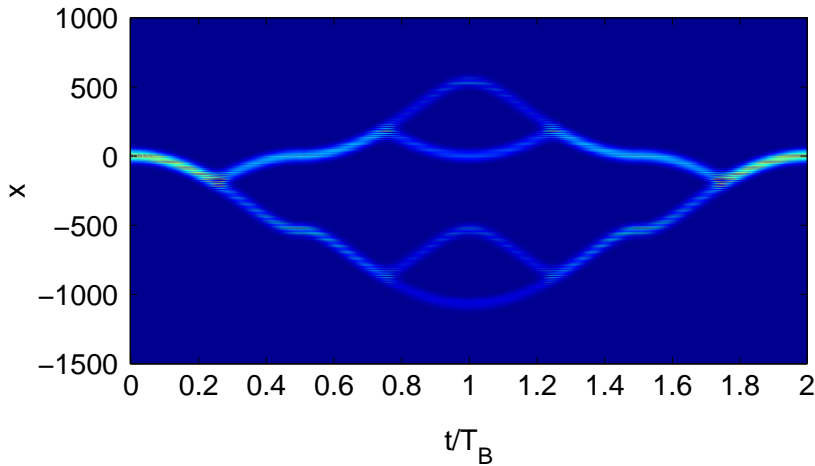}
\caption{\label{5_28} 
Splitting of a gaussian wave packet in position space. Shown is $|\Psi(x,t)|^2$ as a colormap plot for $\varepsilon=0.0825$. The parameter sequence can be found in table 5.}
\end{figure}

Finally we want to point out that the splitting of a wave packet can be easily extended by repeated splitting (see figure \ref{5_28}). Thus even more complex interferometers could be realized. The parameter sequence for the variation of $F$ in figure \ref{5_28} is given in table 5.
\begin{table}[h]
\centering
\begin{tabular}{|c  | c c c | c c c | c c c | c c c| }
\hline
$t/T_B$ &   \multicolumn{2}{l}{0} 
        &   \multicolumn{2}{c}{0.5} 
	& & \multicolumn{2}{c}{1} 
	& & \multicolumn{2}{c}{1.5} 
	&   \multicolumn{2}{r|}{2}  \\ \hline
$F$           &&   $+$  & & &  $-$  & & &  $-$  & &&   $+$  & \\ \hline
\end{tabular}
\caption{}
\end{table}

Mach-Zehnder interferometry by repeated Landau-Zener tunneling in the energy domain was previously discussed for different systems (see, e.g., \cite{Shim91, Lubi90, Sill05}). In contrast, Bloch-Zener oscillations also lead to a {\it spatial} separation of the two branches, which resembles much closer the original Mach-Zehnder setup. 

\subsection{Probing small nonlinearities}
\label{probing}
During the last years, the number of experiments investigating the dynamics of Bose-Einstein condensates (BECs) in optical lattices increased rapidly.
For very low temperatures, the dynamics of a BEC is well described
by the Gross-Pitaevskii or nonlinear Schr\"odinger equation (see, e.g. \cite{Pita03}).
Here, the interactions of the atoms in the condensate are taken into account
in a mean-field approach, that gives rise to an additional potential proportional
to the condensate density, which leads to the Gross-Pitaevskii equation
\be
\fl
\hspace{1.8cm}
  \quad \rmi\hbar \frac{\partial}{\partial t} \Psi(x,t)=
  \eka-\frac{\hbar^2}{2}\frac{\partial^2}{\partial x^2}
  + \cos x + \varepsilon \, \cos \frac{x}{2} + F x
  + g|\Psi(x,t)|^2 \ekz \Psi(x,t)  \, .
\ee
The wave function is normalized to $\| \Psi \|=1$.
Thus, from the mathematical point of view, the difference to the single-particle Hamiltonian is the nonlinear mean-field potential $g|\Psi(x,t)|^2$.
It was shown that a weak nonlinear interaction leads to damping and revival
phenomena of Bloch oscillations \cite{04bloch_bec,Trom01}
while a stronger interaction destroys this coherence effects
immediately \cite{Wu03,Fall04}.

Here we will demonstrate that Bloch-Zener oscillations can be used to probe
very weak nonlinearities, the effect of which would be negligible otherwise.
This can be achieved by the Mach-Zehnder interferometer
setup introduced in the preceding section. Here the phase shift
in both branches of the interferometer is caused by the nonlinear mean-field potential.
A difference of the condensate density in both branches will lead
to a phase shift of both wave packets depending on the interaction
constant $g$. Thus, the condensate density in the two output branches
of the interferometer will also vary with $g$.

\begin{figure}[htb]
\centering
\includegraphics[height=6cm,  angle=0]{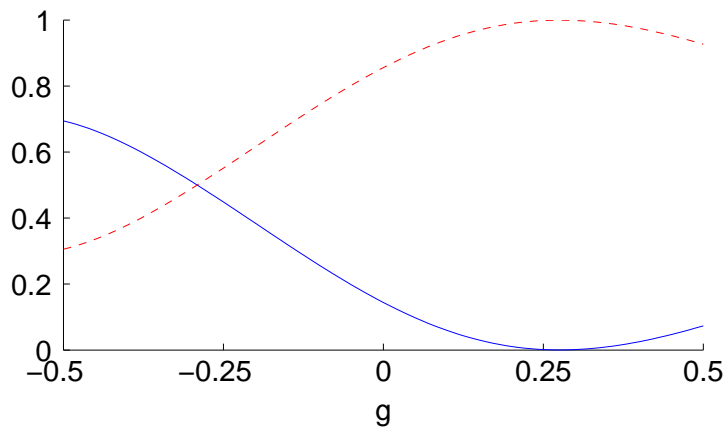}
\caption{\label{Z_2}
Occupation in the output branches of a nonlinear Mach-Zehnder interferometer.
The integral of $|\Psi(x,T_B)|^2$ once over a range of
$-800\leq x\leq-300$ (lower output branch, solid blue line) and once over a range
of $-300\leq x\leq 200$ (upper output branch, dashed red line) are plotted in dependence
of the interaction strength $g$ for $\varepsilon=0.104$.}
\end{figure}

To analyze this effect the time evolution of a normalized gaussian
wave packet
\be
  \Psi(x,t=0) \sim \rme^{-\rka\frac{x}{20}\rkz^2}
\ee
over one Bloch period was calculated numerically for $\varepsilon=0.104$
and different values of the interaction strength $g$.
The resulting density in both output branches of the interferometer calculated
by intergrating $|\Psi(x,T_B)|^2$ over the respective spatial intervals
(cf. figure \ref{5_24}) are plotted in figure \ref{Z_2}.
One observes that the output depends strongly on the
interaction strength even for quite small values of $g$.
For the considered parameter values nonlinear damping or dephasing effects 
are still negligible \cite{04bloch_bec}. The only noticeable effect
of the nonlinearity is the influence on the interferometer output.
Therefore it should be possible to probe very weak interactions with
an interferometric setup as described here.

\section{Conclusion and outlook}
In conclusion, we have investigated the possibility to engineer the
dynamics of matter waves in periodic potentials. A wave packet
in a periodic potential under the influence of a static field performs
an oscillatory motion, the Bloch oscillation. Directed quasi dispersion-free transport can be
realized by a periodic field flip. 

Introducing a weak additional, double-periodic potential offers even
richer opportunities. The Bloch bands split into two minibands, so that
the interplay between Zener tunneling and Bloch oscillations, denoted as
Bloch-Zener oscillations, becomes important. Tunneling between the
minibands leads to a splitting of a wave packet and interference
phenomena.  Combined with the shuttling transport mechanism, one can
implement highly controllable matter wave beam splitters. Furthermore,
Bloch-Zener oscillations provide a natural Mach-Zehnder interferometer
for matter waves. This interferometer can be used, e.g., to probe weak
nonlinear mean-field interactions in Bose-Einstein condensates.

The techniques described in this paper should be experimetally
realizable without major problems. A possible application of the
Mach-Zehnder interferometer could be the detection of the mean-field
interaction in an atom laser beam. 

Up to now we considered only weak external fields $F$, for which decay
is negligible. Since Wannier-Stark eigenstates are resonance states
\cite{02wsrep}, the eigenenergies (the Wannier-Stark ladders) 
are generally complex, i.e. we obtain decay for strong fields. 
Decay rates for the nonlinear Wannier-Stark problem were first calculated
only recently \cite{Wimb05,Wimb06}.
The role of decay is a bit more involved in double-periodic
potentials since the dynamics usually takes place in two miniladders instead of a single one.
The splitting, altering and shifting of resonant tunneling peaks of the decay rate for the two miniladders will be discussed in more detail in a future publication.

\section*{Appendix}
\subsection*{Proof that the dispersion of a gaussian wave packet is proportional to ${t^2}/{\sigma^4}$}
This result can be proven rigorously in a single-band tight-binding
approximation. Here, the Hamiltonian reads
\be
  \hat H(t) = -\frac{\Delta}{4} \sum_n \left( |n+1\rangle\langle n| + |n-1\rangle\langle n|  \right)
  + dF(t) \sum_n n|n\rangle\langle n|\, ,
\ee
where $|n\rangle$ is the Wannier state located at the $n$-th lattice position.
The inital state in the Wannier basis
\be
  |\Psi(0)\rangle = \sum_n c_n |n\rangle
\ee
is assumed to be gaussian, i.e.
\be
  c_n \sim  \re^{-n^2/4\sigma_n^2} \, .
\ee
The time evolution of the position expectation value of the wave packet and its square
can be calculated conveniently using the Lie-algebraic approach introduced in \cite{03TBalg}.
If the inital state is symmetric with respect to the origin and the coefficients $c_n$ are real, as assumed
throughout this section, the results from \cite{03TBalg} simplify and one finds,
\begin{eqnarray}
  \langle \hat N \rangle_t &=& 2 K |\chi_t| \sin(\phi_t)
  \quad \mbox{and} \nn \\
  \langle \hat N^2 \rangle_t &=&  \langle \hat N^2 \rangle_0
  + 2 |\chi_t|^2( 1-L\cos(2\phi_t) )
\end{eqnarray}
with the position operator
\be
\hat N = \sum_n n |n\rangle\langle n |
\ee
and with the coefficients
\be
  \eta_t = \int_0^t \frac{dF(\tau)}{\hbar} \rd \tau \qquad \mbox{and} \qquad
  \chi_t = -\frac{\Delta}{4\hbar} \int_0^t \re^{-\ri \eta_\tau} \rd \tau =: |\chi_t| \re^{-\ri \phi_t}\, .
\ee
The time evolution depends on the initial state through the coherence parameters
\be
  K = \sum_n c_{n-1} c_n \qquad \mbox{and} \qquad   L = \sum_n c_{n-2} c_n \, .
  \label{eqn-coh-par}
\ee
The dispersion of a wave packet 
subjected to the Bloch shuttle $F(t) = \pm F_0$
is given by the increase of the width
\be
\Delta_N^2(t)=\langle\hat N^2\rangle_t-\langle \hat N \rangle ^2_t
\ee
of the wave packet,
\be
  \Delta_N^2(nT_B) - \Delta_N^2(0) = 2 n^2|\chi_{T_B}|^2 (1+L-2K^2) \, .
\ee
For a broad inital state one can replace the sums in (\ref{eqn-coh-par}) by integrals which
yields
\begin{eqnarray}
  K &=& \frac{\int \re^{-(n-1)^2/4\sigma_n^2} \re^{-n^2/4\sigma_n^2} \rd n}{\int \re^{-n^2/2\sigma_n^2} \rd n}
  = \re^{-1/8\sigma_n^2} \, , \nn \\
  L &=& \frac{\int \re^{-(n-2)^2/4\sigma_n^2} \re^{-n^2/4\sigma_n^2} \rd n}{\int \re^{-n^2/2\sigma_n^2} \rd n}
  = \re^{-1/2\sigma_n^2} \, .
\end{eqnarray}
Thus the dispersion is given by
\bea
  \Delta_N^2(nT_B) - \Delta_N^2(0) &= \frac{2 \Delta^2}{F_0^2d^2}
  \left( 1 + \re^{-1/2\sigma_n^2} - 2  \re^{-1/4\sigma_n^2} \right)n^2\nn\\ 
  &=\frac{\Delta^2}{8 F_0^2d^2 \sigma_n^4}\,n^2 + \mathcal{O}(\sigma_n^{-6})\,n^2 \, ,
\eea
where $\chi_{T_B} = \ri \frac{\Delta}{F_0d}$ has been inserted. Since we still have to change from the Wannier basis to the position space we set
\be
\sigma_x\approx d\sigma_n \quad \mbox{and} \quad \Delta_x(t)\approx d \Delta_N(t) \,.   
\ee
Thus the dispersion is in leading order given by 
\be
\Delta_x^2(nT_B) - \Delta_x^2(0) \approx \frac{\Delta^2d^4}{8 F_0^2}\,\frac{n^2}{\sigma_x^4} \, .
\ee

\section*{Acknowledgements}
Support from the Deutsche Forschungsgemeinschaft
via the Graduiertenkolleg `Nichtlineare Optik und Ultrakurzzeitphysik'
and the Studienstiftung des deutschen Volkes
is gratefully acknowledged.
We would also like to thank J. R. Anglin for stimulating discussions.

\section*{References}

\bibliographystyle{unsrtot}

\begin{thebibliography}{10}

\bibitem{Eier03}
B.~Eiermann, P.~Treutlein, Th. Anker, M.~Albiez, M.~Taglieber, K.-P. Marzlin,
  and M.~K. Oberthaler,  Phys. Rev. Lett.  {\bf 91}  (2003)   060402

\bibitem{Mors06}
O.~Morsch and M.~Oberthaler,  Rev. Mod. Phys.  {\bf 78}  (2006)   179

\bibitem{Grei02}
M.~Greiner, O.~Mandel, T.~Esslinger, T.~W. H\"ansch, and I.~Bloch,  Nature
  {\bf 415}  (2002)   39

\bibitem{Daha96}
M.~Ben Dahan, E.~Peik, J.~Reichel, Y.~Castin, and C.~Salomon,  Phys. Rev. Lett.
   {\bf 76}  (1996)   4508

\bibitem{Peik97}
E.~Peik, M.~B. Dahan, I.~Bouchoule, Y.~Castin, and C.~Salomon,  Phys. Rev. A
  {\bf 55}  (1997)   2989

\bibitem{Ande98}
B.~P. Anderson and M.~A. Kasevich,  Science  {\bf 282}  (1998)   1686

\bibitem{02pulse}
M.~Gl{\"u}ck, F.~Keck, and H.~J. Korsch,  Phys. Rev. A  {\bf 66}  (2002)
  043418

\bibitem{Cris04}
M.~Cristiano, O.~Morsch, N.~Malossi, M.~Jona-Lasinio, M.~Anderlini,
  E.~Courtade, and E.~Arimondo,  Optics Express  {\bf 12}  (2004)   4

\bibitem{Bloc28}
F.~Bloch,  Z. Phys  {\bf 52}  (1928)   555

\bibitem{04bloch1d}
T.~Hartmann, F.~Keck, H.~J. Korsch, and S.~Mossmann,  New J. Phys.  {\bf 6}
  (2004)   2

\bibitem{04bloch2d}
D.~Witthaut, F.~Keck, H.~J. Korsch, and S.~Mossmann,  New J. Phys.  {\bf 6}
  (2004)   41

\bibitem{04bloch}
A.~R. Kolovsky and H.~J. Korsch,  Int. J. Mod. Phys. B  {\bf 18}  (2004)   1235

\bibitem{03bloch2D}
A.~R. Kolovsky and H.~J. Korsch,  Phys. Rev. A  {\bf 67}  (2003)   063601

\bibitem{Avro77}
J.~E. Avron, J.~Zak, A.~Grossmann, and L.~Gunther,  J. Math. Phys.  {\bf 18}
  (1977)   918

\bibitem{02wsrep}
M.~Gl{\"u}ck, A.~R. Kolovsky, and H.~J. Korsch,  Phys. Rep.  {\bf 366}  (2002)
   103

\bibitem{03TBalg}
H.~J. Korsch and S.~Mossmann,  Phys. Lett. A  {\bf 317}  (2003)   54

\bibitem{Gore98}
L.~Y. Gorelik, A.~Isacsson, M.~V. Voinova, B.~Kasemo, R.~I. Shekhter, and
  M.~Jonson,  Phys. Rev. Lett.  {\bf 80}  (1998)   4526

\bibitem{06bloch_zener}
B.~M. Breid, D.~Witthaut, and H.~J. Korsch,  New J. Phys.  {\bf 8}
  (2006)   110

\bibitem{Feit82}
M.~D. Feit, J.~A.~Fleck Jr., and A.~Steiger,  J. Comput. Phys.  {\bf 47}
  (1982)   412

\bibitem{Land32}
L.~D. Landau,  Phys. Z. Sowjet.  {\bf 1}  (1932)   88

\bibitem{Zene32}
C.~Zener,  Proc. Roy. Soc. Lond. A  {\bf 137}  (1932)   696

\bibitem{Shim91}
E.~Shimshoni and Y.~Gefen,  Ann. Phys.  {\bf 210}  (1991)   16

\bibitem{Lubi90}
D.~Lubin, Y.~Gefen, and I.~Goldhirsch,  Physica A  {\bf 168}  (1990)   456

\bibitem{Sill05}
M.~Sillanp\"a\"a, T.~Lehtinen, A.~Paila, Y.~Makhlin, and P.~Hakonen,
  cond-mat/0510559  (2005)  

\bibitem{Pita03}
L.~Pitaevskii and S.~Stringari,  {\em Bose-Einstein Condensation},   Oxford
  University Press, Oxford, 2003

\bibitem{04bloch_bec}
D.~Witthaut, M.~Werder, S.~Mossmann, and H.~J. Korsch,  Phys. Rev. E  {\bf 71}
  (2005)   036625

\bibitem{Trom01}
A.~Trombettoni and A.~Smerzi,  Phys. Rev. Lett.  {\bf 86}  (2001)   2353

\bibitem{Wu03}
Biao Wu and Qian Niu,  New J. Phys.  {\bf 5}  (2003)   104

\bibitem{Fall04}
L.~Fallani, L.~De Sarlo, J.~E. Lye, M.~Modugno, R.~Sears, C.~Fort, and
  M.~Inguscio,  Phys. Rev. Lett.  {\bf 93}  (2004)   140406

\bibitem{Wimb05}
S.~Wimberger, R.~Mannella, O.~Morsch, E.~Arimondo, A.~R. Kolovsky, and
  A.~Buchleitner,  Phys. Rev. A  {\bf 72}  (2005)   063610

\bibitem{Wimb06}
S.~Wimberger, P.~Schlagheck, and R.~Mannella,  J. Phys. B  {\bf 39}  (2006)
  729

\end{thebibliography}

\end{document}